\documentclass{article}

\usepackage[utf8]{inputenc}
\usepackage{tismir}
\usepackage{amsmath}
\usepackage{hyperref}
\usepackage{url}
\usepackage{booktabs}
\usepackage{graphicx}
\usepackage{color}
\usepackage{booktabs}
\usepackage{lipsum}

\title{GiantMIDI-Piano: A large-scale MIDI Dataset for Classical Piano Music}
\author{%
Qiuqiang Kong\thanks{ByteDance}, Bochen Li\protect\footnotemark[1], Jitong Chen\protect\footnotemark[1], Yuxuan Wang\protect\footnotemark[1]}
\date{}

\authorref{Anonymous}
\authorshort{Anonymous}
\begin{document}

%%%%%%%%%%%%%%%%%%%%%%%%%%%%%%%%%%%%%%%%%%%%%%%%%%%%%%%%%%%%%%%%%%%%%%%%%%%%%%%%
% Abstract
%%%%%%%%%%%%%%%%%%%%%%%%%%%%%%%%%%%%%%%%%%%%%%%%%%%%%%%%%%%%%%%%%%%%%%%%%%%%%%%%

\twocolumn[{%
\maketitleblock
\begin{abstract}
Symbolic music datasets are important for music information retrieval and musical analysis. However, there is a lack of large-scale symbolic datasets for classical piano music. In this article, we create a GiantMIDI-Piano (GP) dataset containing 38,700,838 transcribed notes and 10,855 unique solo piano works composed by 2,786 composers. We extract the names of music works and the names of composers from the International Music Score Library Project (IMSLP). We search and download their corresponding audio recordings from the internet. We further create a curated subset containing 7,236 works composed by 1,787 composers by constraining the titles of downloaded audio recordings containing the surnames of composers. We apply a convolutional neural network to detect solo piano works. Then, we transcribe those solo piano recordings into Musical Instrument Digital Interface (MIDI) files using a high-resolution piano transcription system. Each transcribed MIDI file contains the onset, offset, pitch, and velocity attributes of piano notes and pedals. GiantMIDI-Piano includes 90\% live performance MIDI files and 10\% sequence input MIDI files. We analyse the statistics of GiantMIDI-Piano and show pitch class, interval, trichord, and tetrachord frequencies of six composers from different eras to show that GiantMIDI-Piano can be used for musical analysis. We evaluate the quality of GiantMIDI-Piano in terms of solo piano detection F1 scores, metadata accuracy, and transcription error rates. We release the source code for acquiring the GiantMIDI-Piano dataset at \url{https://github.com/bytedance/GiantMIDI-Piano}

\end{abstract}
}]
\saythanks{}

%%%%%%%%%%%%%%%%%%%%%%%%%%%%%%%%%%%%%%%%%%%%%%%%%%%%%%%%%%%%%%%%%%%%%%%%%%%%%%%%
% Main Content Start
%%%%%%%%%%%%%%%%%%%%%%%%%%%%%%%%%%%%%%%%%%%%%%%%%%%%%%%%%%%%%%%%%%%%%%%%%%%%%%%%

\section{Introduction}\label{sec:introduction}

Symbolic music datasets are important for music information retrieval (MIR) and musical analysis. In the Western music tradition, musicians use musical notation to write music. This notation includes pitches, rhythms, and chords of music. Musicologists used to analyse music works by reading music notation. Recently, computers have been used to process and analyse large-scale data and have been widely used in MIR. However, there is a lack of large-scale symbolic music datasets covering a wide range of solo piano works.

One difficulty of computer-based MIR is that musical notation such as staffs is not directly readable by a computer. Therefore, converting music notation into computer-readable formats is important. Early works of converting music into symbolic representations can be traced back to the 1900s, when piano rolls \citep{bryner2002piano, shi2019supra} were developed to record music that could be played on a musical instrument. Piano rolls are continuous rolls of paper with perforations punched into them. In 1981, Musical Instrument Digital Interface (MIDI) \citep{smith1981usi} was proposed as a technical standard to represent music and can be read by a computer. MIDI files use event messages to specify the instructions of music, including pitch, onset, offset, and velocity of notes. MIDI files also carry rich information of music events such as sustain pedals. The MIDI format has been popular for music production in recent years.

In this work, we focus on building a large-scale MIDI dataset for classical solo piano music. There are several previous piano MIDI datasets including the piano-midi.de \citep{pianomidide} dataset, the MAESTRO dataset \citep{hawthorne2018enabling}, the classical archives \citep{classicalarchives} dataset, and the Kunstderfuge dataset \citep{kunstderfuge}. However, those datasets are limited to hundreds of composers and hundreds of hours of unique works \citep{kunstderfuge}. MusicXML \citep{good2001musicxml} is another symbolic format of music, while there are fewer MusicXML datasets than MIDI datasets. Other machine-readable formats include the music encoding initiative (MEI) \citep{MEI}, Humdrum \cite{sapp2005online}, and LilyPond (\cite{nienhuys2003lilypond}). Optical music recognition (OMR) \citep{rebelo2012optical, bainbridge2001challenge} is a technique to transcribe image scores into symbolic formats. However, the performance of OMR systems is limited to score qualities.

In this article, we collect and transcribe a large-scale classical piano MIDI dataset called GiantMIDI-Piano. To our knowledge, GiantMIDI-Piano is the largest piano MIDI dataset so far. GiantMIDI-Piano is collected as follows: 1) We parse the names of composers and the names of music works from the International Music Score Library Project (IMSLP)\footnote{\url{https://imslp.org}}; 2) We search and download audio recordings of all matching music works from YouTube; 3) We build a solo piano detection system to detect solo piano recordings; 4) We transcribe solo piano recordings into MIDI files using a high-resolution piano transcription system \citep{kong2020high}. In this article, we analyse the statistics of GiantMIDI-Piano, including the number of works, durations of works, and nationalities of composers. In addition, we analyse the statistics of note, interval, and chord distribution of six composers from different eras to show that GiantMIDI-Piano can be used for musical analysis.

\subsection{Applications}
The GiantMIDI-Piano dataset can be used in many research areas, including 1) Computer-based musical analysis \citep{volk2011unfolding, meredith2016computational} such as using computers to analyse the structure, chords, and melody of music works. 2) Symbolic music generation \citep{yang2017midinet, hawthorne2018enabling} such as generating symbolic music in symbolic format. 3) Computer-based music information retrieval \citep{casey2008content, choi2017tutorial} such as music transcription and music tagging. 4) Expressive performance analysis \citep{cancino2018computational} such as analysing the performance of different pianists.

This paper is organised as follows: Section 2 surveys piano MIDI datasets; Section 3 introduces the collection of the GiantMIDI-Piano dataset; Section 4 investigates the statistics of the GiantMIDI-Piano dataset; Section 5 evaluates the quality of the GiantMIDI-Piano dataset.

\section{Dataset Survey}
We introduce several piano MIDI datasets as follows. The Piano-midi.de dataset \citep{pianomidide} contains classical solo piano works entered by a MIDI sequencer. Piano-midi.de contains 571 works composed by 26 composers with a total duration of 36.7 hours till Feb. 2020. The classical archives collection \citep{classicalarchives} contains a large number of MIDI files of classical music, including both piano and non-piano works. There are 133 composers with a total duration of 46.3 hours of MIDI files in this dataset. The KernScores dataset \citep{sapp2005online} contains classical music with a Humdrum format and is obtained by an optical music recognition system. The Kunstderfuge dataset \citep{kunstderfuge} contains solo piano and non-solo piano works of 598 composers. All of the piano-midi.de, classical archives, and Kunstderfuge datasets are entered using a MIDI sequencer and are not played by pianists. 

The MAPS dataset \citep{emiya2010maps} used MIDI files from Piano-midi.de to render real recordings by playing back the MIDI files on a Yamaha Disklavier. The MAESTRO dataset \citep{hawthorne2018enabling} contains over 200 hours of fine alignment MIDI files and audio recordings. In MAESTRO, virtuoso pianists performed on Yamaha Disklaviers with an integrated MIDI capture system. MAESTRO contains music works from 62 composers. There are several duplicated works in MAESTRO. For example, there are 11 versions of \textit{Scherzo No. 2 in B-flat Minor, Op. 31} composed by Chopin. All duplicated works are removed when calculating the number and duration of works. 

\begin{table}
\footnotesize
\centering
\caption{Piano MIDI datasets. GP is the abbreviation for GiantMIDI-Piano.}
\label{table:dataset_survey}
\begin{tabular}{lcccc}
 \toprule
 Dataset & Composers & Works & Hours & Type \\
 \midrule
 piano-midi.de & 26 & 571 & 37 & Seq. \\
 Classical archives & 133 & 856 & 46 & Seq. \\
 Kunstderfuge & 598 & - & - & Seq. \\
 KerbScores & - & - & - & Seq. \\
 SUPRA & 111 & 410 & - & Perf. \\
 ASAP & 16 & 222 & - & Perf. \\
 MAESTRO & 62 & 529 & 84 & Perf. \\
 MAPS & - & 270 & 19 & Perf. \\
 \textbf{GiantMIDI-Piano} & \textbf{2,786} & \textbf{10,855} & \textbf{1,237} & \textbf{90\% Perf.} \\
 \textbf{Curated GP} & \textbf{1,787} & \textbf{7,236} & 875 & \textbf{89\% Perf.} \\
 \bottomrule
\end{tabular}
\end{table}

Table \ref{table:dataset_survey} shows the number of composers, the number of unique works, total durations, and data types of different MIDI datasets. Data types include sequenced (Seq.) MIDI files input by MIDI sequencers and performed (Perf.) MIDI files played by pianists. There are other MIDI datasets including the Lakh dataset \citep{raffel2016learning}, the Bach Doodle dataset \citep{bachdoodle2019}, the Bach Chorales dataset \citep{conklin1995multiple}, the URMP dataset \citep{li2018creating}, the Bach10 dataset \citep{duan2010multiple}, the CrestMusePEDB dataset \citep{hashida2008new}, the SUPRA dataset \citep{shi2019supra}, and the ASAP dataset \citep{foscarin2020asap}. \citep{huang2018music} collected 10,000 hours of piano recordings for music generation, but the dataset is not publicly available.

\section{GiantMIDI-Piano Dataset}

\subsection{Metadata from IMSLP}
To begin with, we acquire the names of composers and the names of music works by parsing the webpages of the International Music Score Library Project \citep{imslp}, the largest publicly available music library in the world. In IMSLP, each composer has a webpage containing the list of their work names. We acquire 143,701 music works composed by 18,067 composers by parsing those web pages. For each composer, if there exists a biography link on the composer page, we access that biography link to search for their birth year, death year, and nationality. We set the birth year, death year, and nationality to ``unknown'' if a composer does not have such a biography link. We obtain the nationalities of 4,274 composers and births of 5,981 composers out of 18,067 composers by automatically parsing the biography links. 

As the automatically parsed meta information of composers from the internet is incomplete, we manually check the nationalities, births, and deaths for 2,786 composers. We label 2,291 birth years, 2,254 death years, and 2,115 nationalities by searching the information of composers on the internet. We label not found birth years, death years, and nationalities as ``unknown''. We create metadata files containing the information of composers and music works, respectively.

\subsection{Search Audio}
We search audio recordings on YouTube by using a keyword of \textit{first name, surname, music work name} from the metadata. For each keyword, we select the first returned result on YouTube. However, there can be cases where the returned YouTube title may not exactly match the keyword. For example, for a keyword \textit{Frédéric Chopin, Scherzo No.2 Op.31}, the top returned result can be \textit{Chopin - Scherzo No. 2, Op. 31 (Rubinstein)}. Although the keyword and the returned YouTube title are different, they indicate the same music work. We denote the set of words in a searching keyword as $ X $ and the set of words in a returned YouTube title as $ Y $. We propose a modified Jaccard similarity \citep{niwattanakul2013using} to evaluate how much a keyword and a returned result are matched. 

The original Jaccard similarity is defined as $ J = |X \cap Y| / (|X| \cup |Y|) $. The drawback of this original Jaccard similarity is that the length of the searched YouTube title $ |Y| $ can be long, so that $ J $ will be small. This is often the case because searched YouTube titles usually contain extra words such as the names of performers and the dates of performances. Our aim is to define a metric where the denominator only depends on the searching keyword $ |X| $ and is independent of the length of the searched YouTube title $ |Y| $. We propose a modified Jaccard similarity \citep{niwattanakul2013using} between $ X $ and $ Y $ as:
\begin{equation} \label{eq:jaccard}
J = |X \cap Y| / |X|.
\end{equation}
\noindent Higher $ J $ indicates that $ X $ and $ Y $ have larger similarity, and lower $ J $ indicates that $ X $ and $ Y $ have less similarity. We empirically set a similarity threshold to 0.6 to balance the precision and recall of searched results. If $ J $ is strictly larger than this threshold, then we say $ X $ are $ Y $ are matched, and vice versa. In total, we retrieve and download 60,724 audio recordings out of 143,701 music works.

\subsection{Solo Piano Detection}
We detect solo piano works from IMSLP to build the GiantMIDI-Piano dataset. Filtering music works with keywords containing ``piano'' may lead to incorrect results. For example, a ``Piano Concerto'' is an ensemble of piano and orchestra which is not solo piano. On the other hand, the keyword \textit{Chopin, Frédéric, Nocturnes, Op.62} does not contain the word ``piano'', but it is indeed a solo piano. To address this problem, we train an audio-based solo piano detection system using a convolutional neural network (CNN) \citep{kong2020panns}. The piano detection system takes 1-second segments as input and extracts log mel spectrograms as input to the CNN. 

The CNN consists of four convolutional layers. Each convolutional layer consists of a linear convolutional operation with a kernel size of $ 3 \times 3 $, a batch normalization \citep{ioffe2015batch}, and a ReLU nonlinearity \citep{glorot2011deep}. The output of the CNN predicts the solo piano probability of a segment. Binary cross-entropy is used as a loss function to train the CNN. We collect solo piano recordings as positive samples and collect music and other sounds as negative samples. In addition, the mixtures of piano and other sounds are also used as negative samples. In inference, we average the predictions of all 1-second segments of a recording to calculate its solo piano probability. We regard an audio recording as solo piano if the probability is strictly larger than 0.5 and vice versa. In total, we obtain 10,855 solo pianos composed by 2,786 composers out of 60,724 downloaded audio recordings. These 10,855 audio files are transcribed into MIDI files which constitute the full GiantMIDI-Piano dataset.

\begin{figure*}[t]
  \centering
  \centerline{\includegraphics[width=\textwidth]{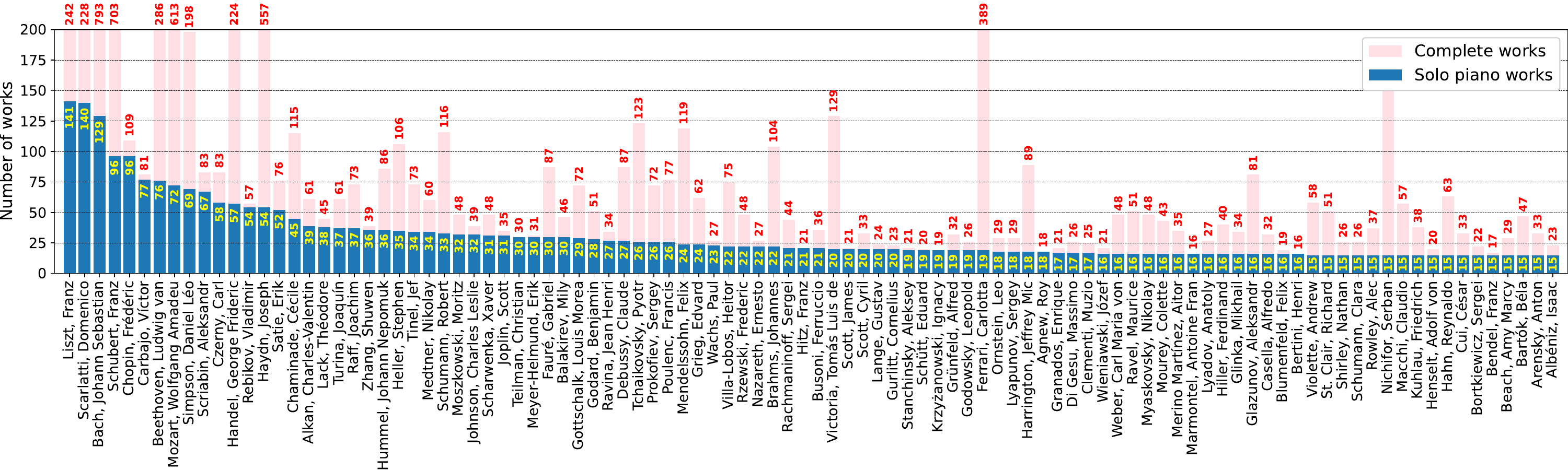}}
  \caption{Number of solo piano works of the curated GP dataset. Top 100 are shown.}
  \label{fig:composer_works_num}
\end{figure*}

\begin{figure*}[t]
  \centering
  \centerline{\includegraphics[width=\textwidth]{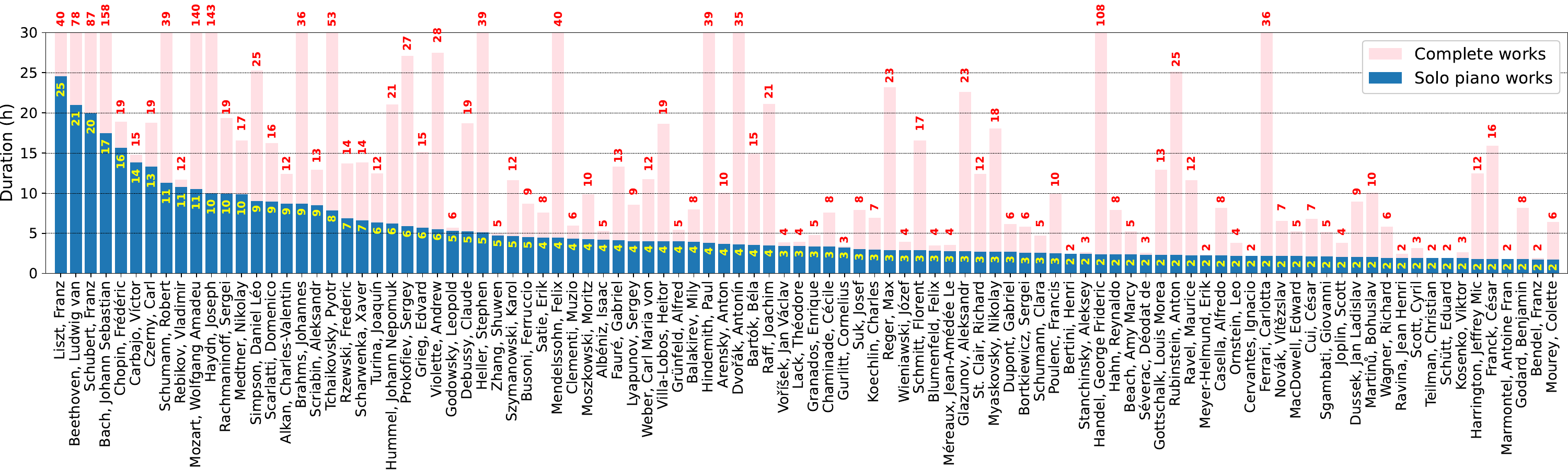}}
  \caption{Duration of solo piano works of the curated GP dataset. Top 100 are shown.}
  \label{fig:composer_durations}
\end{figure*}

\begin{figure}[t]
  \centering
  \centerline{\includegraphics[width=\columnwidth]{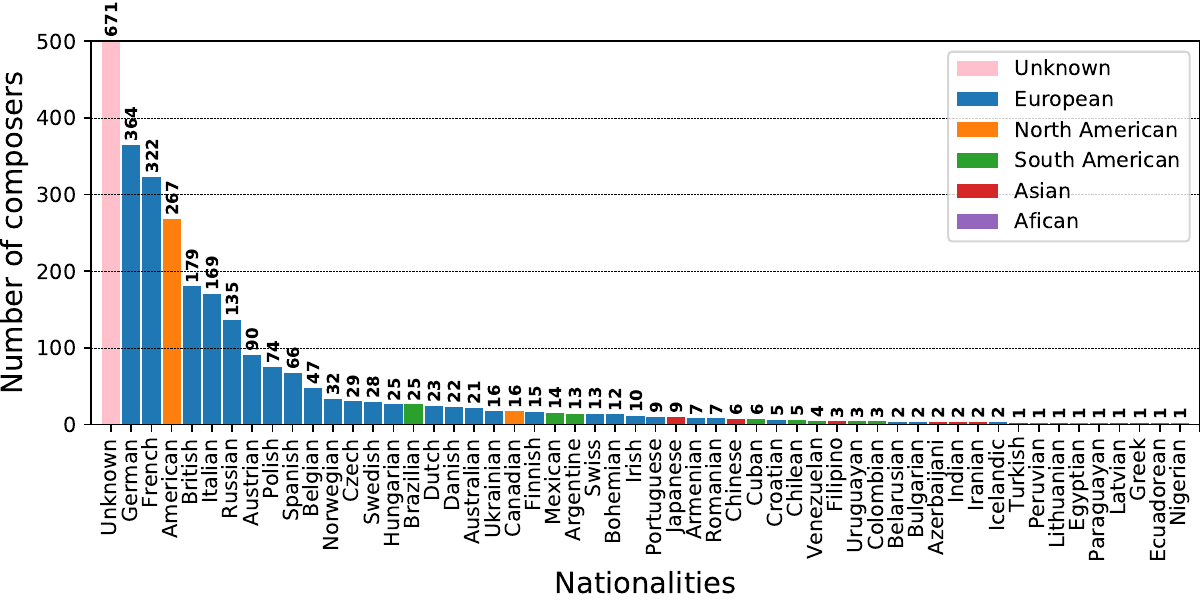}}
  \caption{Number of composers with different nationalities of the full GP dataset.}
  \label{fig:nationalities}
\end{figure}

\subsection{Constrain Composer Surnames}
Among the detected 10,855 solo piano works, there are several music works composed by not well-known composers but are attached to famous composers. For example, there are 273 searched music works assigned to Chopin, but only 102 of them are actually composed by Chopin, while other music works are composed by other composers. To alleviate this problem, we create a curated subset by constraining the titles of downloaded audio recordings containing the surnames of composers. After this constraint, we obtain a curated GiantMIDI-Piano dataset containing 7,236 music works composed by 1,787 composers.

\subsection{Piano Transcription}
We transcribe all 10,855 solo piano recordings into MIDI files using an open-sourced high-resolution piano transcription system\footnote{\url{https://github.com/bytedance/piano\_transcription}} \citep{kong2020high}, an improvement over the onsets and frames piano transcription system \citep{hawthorne2017onsets, hawthorne2018enabling} and other systems \citep{kim2019adversarial, kwon2020polyphonic}. The piano transcription system is trained on the training subset of the MAESTRO dataset version 2.0.0 \citep{hawthorne2018enabling}. The training and testing subset contain 161.3 and 20.5 hours of aligned piano recordings and MIDI files, respectively. The piano transcription system predicts all of the pitch, onset, offset, and velocity attributes of notes. The transcribed results also include sustain pedals. For piano note transcription, our system consists of a frame-wise classification, an onset regression, an offset regression, and a velocity regression sub-module. Each sub-module is modeled by a convolutional recurrent neural network (CRNN) with eight convolutional layers and two bi-directional gated recurrent units (GRU) layers. The output of each module has a dimension of 88, equivalent to the number of notes on a modern piano. 

The pedal transcription system has the same architecture as the note transcription system, except that there is only one output after the CRNN sub-module indicating the onset or offset probabilities of pedals. In inference, all piano recordings are converted into mono with a sampling rate of 16 kHz. We use a short-time Fourier transform (STFT) with a Hann window size 2048 and a hop size 160 to extract spectrograms, so there are 100 frames in a second. Then, mel filter banks with 229 bins are used to extract log mel spectrogram as input feature \citep{hawthorne2018enabling}. The transcription system outputs the predicted probabilities of pitch, onset, offset, and velocity. Finally, those outputs are post-processed into MIDI events.

\begin{figure*}[t]
  \centering
  \centerline{\includegraphics[width=\textwidth]{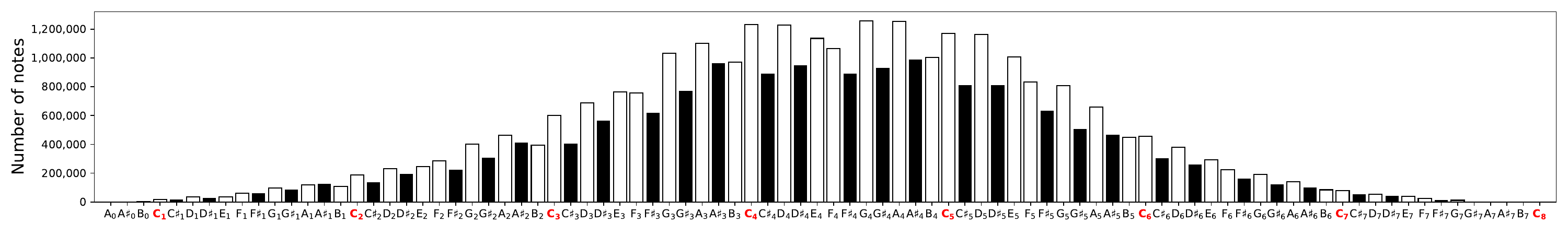}}
  \caption{Note histogram of the curated GP dataset.}
  \label{fig:note_histogram}
\end{figure*}

\begin{figure*}[h!]
  \centering
  \centerline{\includegraphics[width=\textwidth]{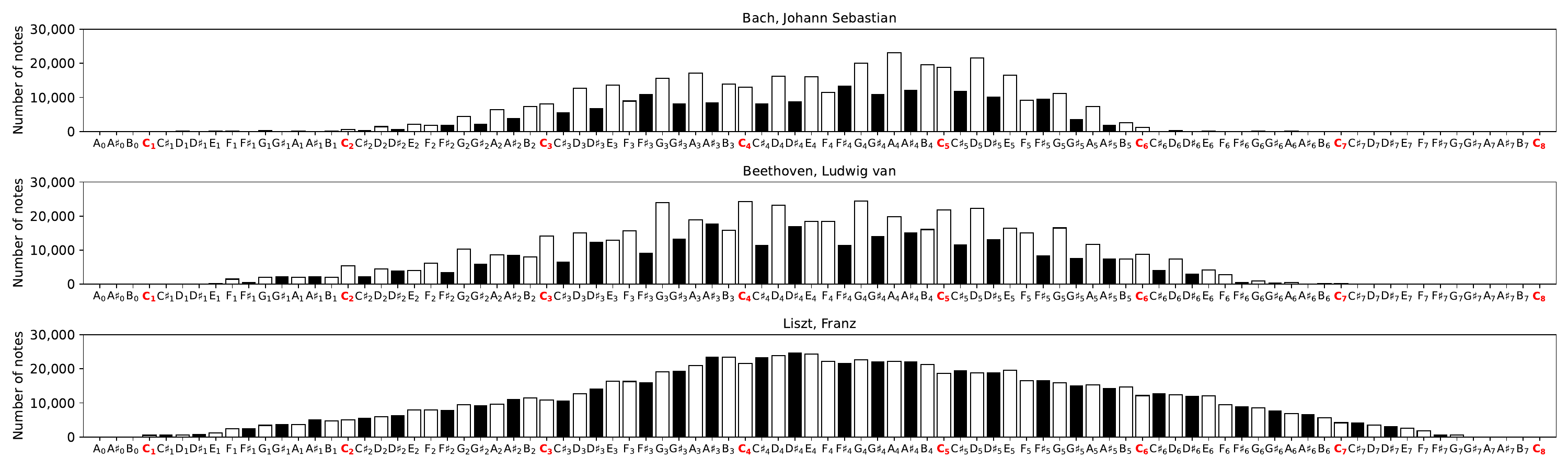}}
  \caption{Notes histogram of J.S. Bach, Beethoven, and Liszt of the curated GP dataset.}
  \label{fig:selected_composers_note_histogram}
\end{figure*}

\begin{figure*}[t]
  \centering
  \centerline{\includegraphics[width=\textwidth]{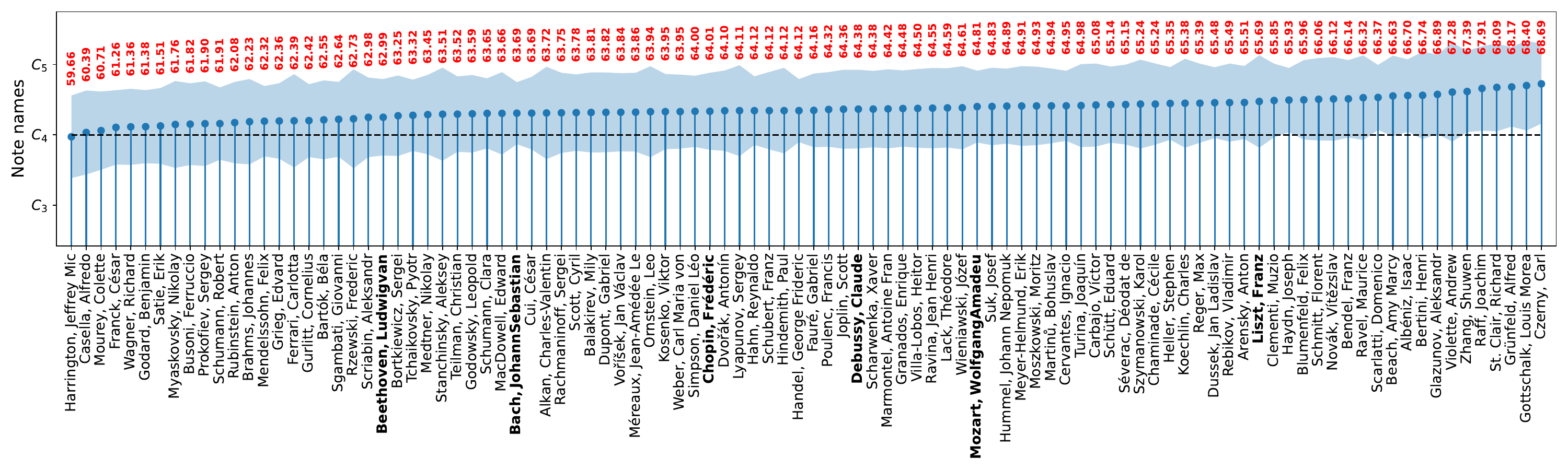}}
  \caption{Pitch distribution of top 100 composers of the curated GP dataset.}
  \label{fig:note_mean_std}
\end{figure*}

\begin{figure*}[t]
  \centering
  \centerline{\includegraphics[width=\textwidth]{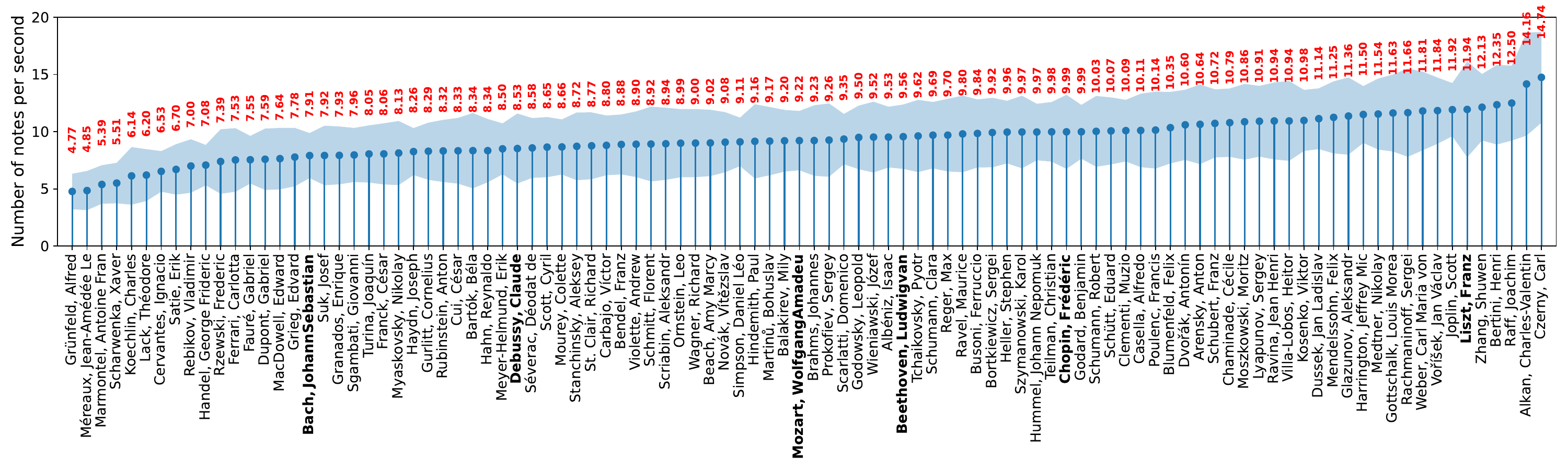}}
  \caption{The number of notes per second of top 100 composers of the curated GP dataset.}
  \label{fig:notes_per_second}
\end{figure*}

\section{Statistics}
We analyse the statistics of GiantMIDI-piano, including the number and duration of music works composed by different composers, the nationality of composers, and the distribution of notes by composers. Then, we investigate the statistics of six composers from different eras by calculating their pitch class, interval, trichord, and tetrachord distributions. All of Fig. \ref{fig:composer_works_num} to Fig. \ref{fig:selected_composers_chords_4} except Fig. \ref{fig:nationalities} are plotted with the statistics of the curated GiantMIDI-Piano dataset. Fig. \ref{fig:nationalities} shows the manually-checked nationalities of 2,786 composers in the full GiantMIDI-Piano dataset.

\subsection{The Number of Solo Piano Works}
Fig. \ref{fig:composer_works_num} shows the numbers of piano works composed by different composers sorted in descending order of the curated GiantMIDI-Piano dataset. Fig. \ref{fig:composer_works_num} shows the statistics of top 100 composers out of 2,786 composers. Blue bars show the number of solo piano works. Pink bars show the number of complete works, including both solo piano and non-solo piano works. Fig. \ref{fig:composer_works_num} shows that there are 141 solo piano works composed by Liszt, followed by 140 and 129 solo piano works composed by Scarlatti and J. S. Bach. Some composers, such as Chopin composed more solo piano works than non-solo piano works. For example, there are 96 solo piano works out of 109 complete works composed by Chopin in the curated GiantMIDI-Piano dataset. Fig. \ref{fig:composer_works_num} shows that the number of solo piano works of different composers has a long tail distribution. 

\subsection{The Duration of Solo Piano Works}\label{section:piano_works_duration}
Fig. \ref{fig:composer_durations} shows the duration of solo piano works composed by different composers sorted in descending order of the curated GiantMIDI-Piano dataset. The duration of works composed by Liszt is the longest at 25 hours, followed by Beethoven at 21 hours and Schubert at 20 hours. Some composers composed more non-piano works than solo piano works. For example, there are 108 hours of complete works composed by Handel in the dataset, while only 2 hours of them are played on a modern piano. The rank of composers in Fig. \ref{fig:composer_durations} is different from Fig. \ref{fig:composer_works_num}, indicating that the average duration of solo piano works composed by different composers are different. 

\subsection{Nationalities of Composers}
Fig. \ref{fig:nationalities} shows the number of composers with different nationalities sorted in descending order of the full GiantMIDI-Piano dataset. The nationality of 2,786 composers are initially obtained from Wikipedia and are later manually checked\footnote{There are debates on the nationality of some composers. We extract the nationality of composers from Wikipedia and do not discuss region debates in this work.}. Fig. \ref{fig:nationalities} shows that there are 671 composers with unknown nationality. There are 364 German composers, followed by 322 French composers and 267 American composers. We color-code the continent of nationalities from ``Unknown'', ``European'', ``North American'', ``South American'', ``Asian'', to ``African''. In GiantMIDI-Piano, the nationalities of most composers are European. The numbers of composers with nationalities from South American, Asian, and African are fewer.

\subsection{Note Histogram}
Fig. \ref{fig:note_histogram} shows the note histogram  of the curated GiantMIDI-Piano dataset. There are 24,253,495 transcribed notes. The horizontal axis shows the scientific pitch notations, which covers 88 notes on a modern piano from $ \text{A}_{0} $ to $ \text{C}_{8} $. Middle C is denoted as $ \text{C}_{4} $. We do not distinguish enharmonic notes, for example, a note $ \text{C}\sharp / \text{D}\flat $ is simply denoted as $ \text{C}\sharp $. The white and black bars correspond to the white and black keys on a modern piano, respectively. Fig. \ref{fig:note_histogram} shows that the note histogram has a normal distribution. The most played note is $ \text{G}_{4} $. There are more notes close to $ \text{G}_{4} $ and less notes far from $ \text{G}_{4} $. The most played notes are within the octave between $ \text{C}_{4} $ and $ \text{C}_{5} $. White keys are being played more often than black keys.

Fig. \ref{fig:selected_composers_note_histogram} visualizes the note histogram of three composers from different eras, including J. S. Bach, Beethoven, and Liszt. The note range of J. S. Bach is mostly between $ \text{C}_{2} $ and $ \text{C}_{6} $ covering four octaves, which is consistent with the note range of a conventional harpsichord or organ. The note range of Beethoven is mostly between $ \text{F}_{1} $ and $ \text{C}_{7} $ covering five and a half octaves. The note range of Liszt is the widest, covering the whole range of a modern piano.

\subsection{Pitch Distribution of Top 100 Composers}
Fig. \ref{fig:note_mean_std} shows the pitch distribution sorted in ascending order over the top 100 composers in Fig. \ref{fig:composer_durations} of the curated GiantMIDI-Piano dataset. The average pitches of most composers are between $ \text{C}_{4} $ and $ \text{C}_{5} $, where $ \text{C}_{4} $ corresponds to a MIDI pitch value 60. The shades indicate the one standard deviation area of pitch distributions. Jeffrey Michael Harrington has the lowest average pitch value of $ \text{C}_{4} $. Carl Czerny has the highest average pitch value of $ \text{A}_{4} $. 

\subsection{The Number of Notes Per Second Distribution of Top 100 Composers}
Fig. \ref{fig:notes_per_second} shows the number of notes per second distribution sorted in ascending order over the top 100 composers in Fig. \ref{fig:composer_durations} of the curated GiantMIDI-Piano dataset. The number of notes per second is calculated by dividing all works notes number by all works duration of a composer. The average numbers of notes per second of most composers are between 5 and 10. The shades indicate the one standard deviation area of the number of notes per second distribution. Alfred Gr\"unfeld has the smallest number of notes per second with a value of 4.18. Carl Czerny has the largest number of notes per second with a value of 13.61.

\subsection{Pitch Class Distribution}\label{section:chroma}
\begin{figure}[t]
  \centering
  \centerline{\includegraphics[width=\columnwidth]{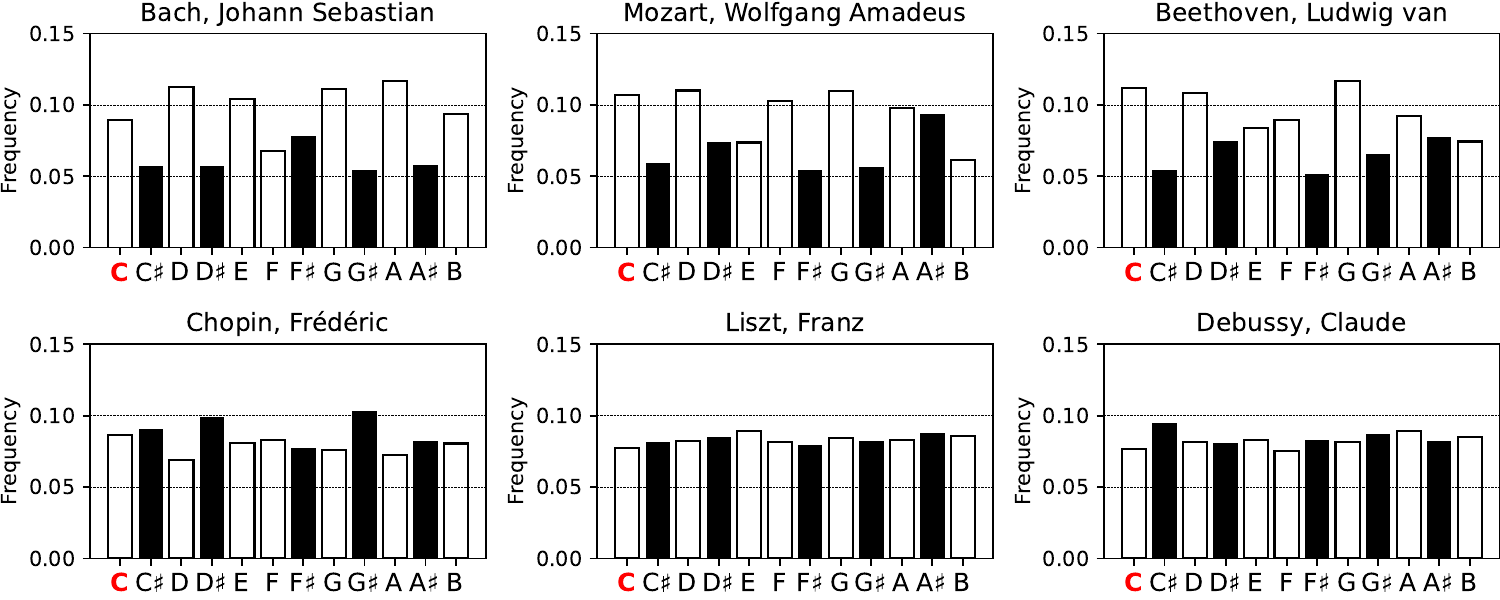}}
  \caption{Pitch class distribution of six composers of the curated GP dataset.}
  \label{fig:selected_composers_chroma}
\end{figure}

We denote the set of pitch names as $ \{ \text{C}, \text{C} \sharp, \text{D}, \text{D}\sharp, \text{E}, \text{F}, \text{F}\sharp, \text{G}, \text{G}\sharp, \text{A}, \text{A}\sharp, \text{B} \} $. The notes from $ \text{C} $ to $ \text{B} $ are denoted as 0 to 11 \citep{forte1973structure}, respectively. We calculate the statistics of six composers from different eras including J. S. Bach, Mozart, Beethoven, Chopin, Liszt, and Debussy. Fig. \ref{fig:selected_composers_chroma} shows that J. S. Bach used $ \text{D},~\text{E},~ \text{G}$, and $\text{A} $ most in his solo piano works. Mozart used $ \text{C},~\text{D},~\text{F}$, and $\text{G} $ most in his solo piano works and used more $ \text{A}\sharp/\text{B}\flat $ than other composers. Beethoven used more $ \text{C},~\text{D}$, and $ \text{G} $ than other notes. Chopin used $ \text{D}\sharp/\text{E}\flat $ and $ \text{G}\sharp/\text{A}\flat $ most in his solo piano works. Liszt and Debussy used all twelve pitch classes more uniformly in their solo piano works than other composers. Liszt used $ \text{E} $ most, and Debussy used $ \text{C}\sharp/\text{D}\flat $ most. As expected, most of Baroque and classical solo piano works were in keys close to $ \text{C} $, whereas Romantic and later composers explored distant keys and tended to use all twelve pitch classes more uniformly.

\begin{figure}[t]
  \centering
  \centerline{\includegraphics[width=\columnwidth]{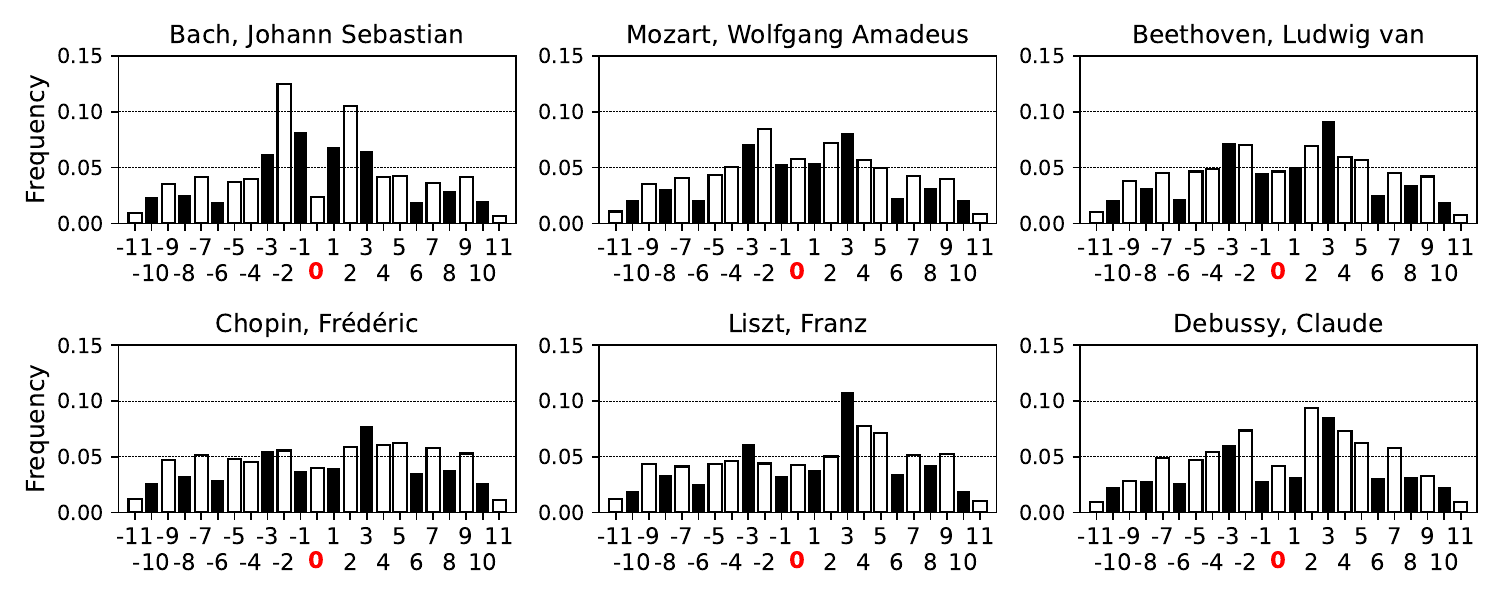}}
  \caption{Interval distribution of six composers of the curated GP dataset.}
  \label{fig:selected_composers_intervals}
\end{figure}

\begin{figure}[t]
  \centering
  \centerline{\includegraphics[width=\columnwidth]{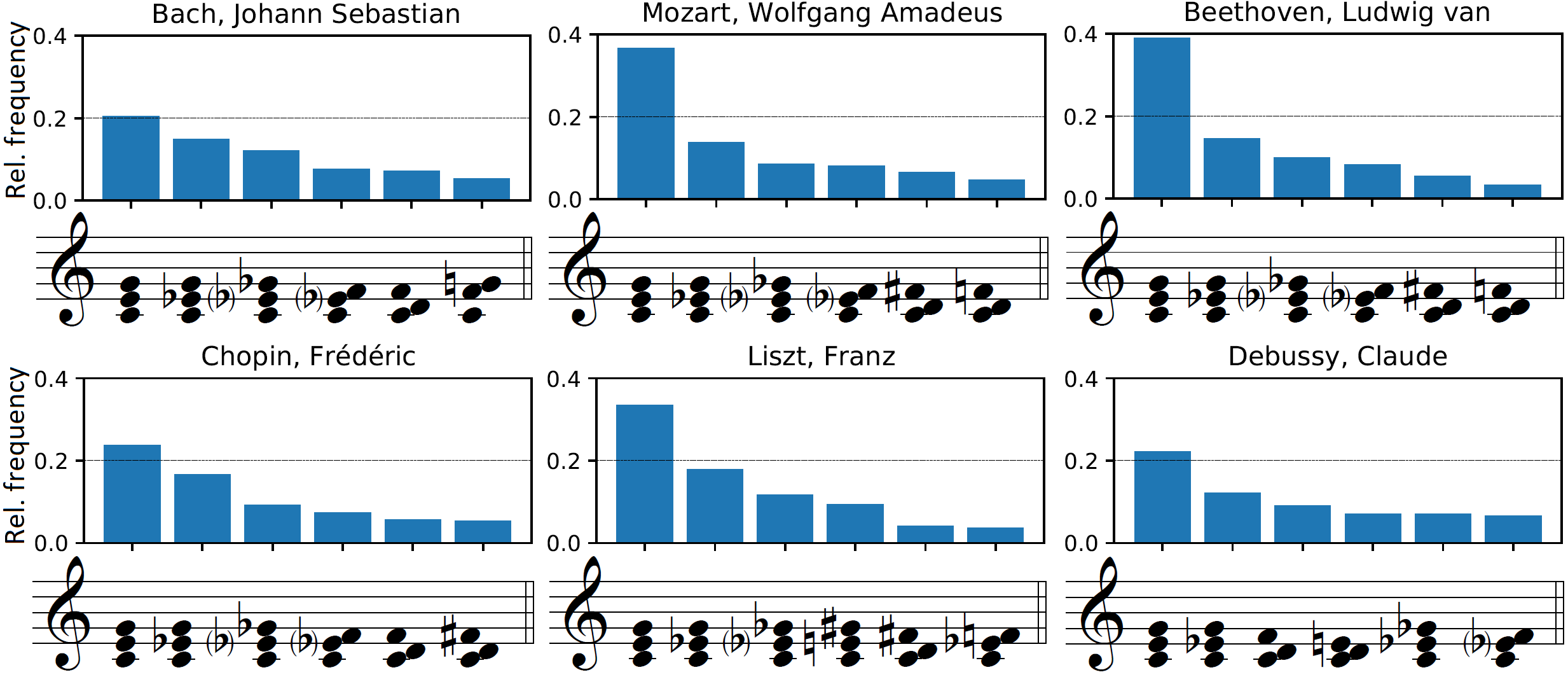}}
  \caption{Trichord distribution of six composers of the curated GP dataset. Rel. frequency is the abbreviation for relative frequency.}
  \label{fig:selected_composers_chords_3}
\end{figure}

\begin{figure}[t]
  \centering
  \centerline{\includegraphics[width=\columnwidth]{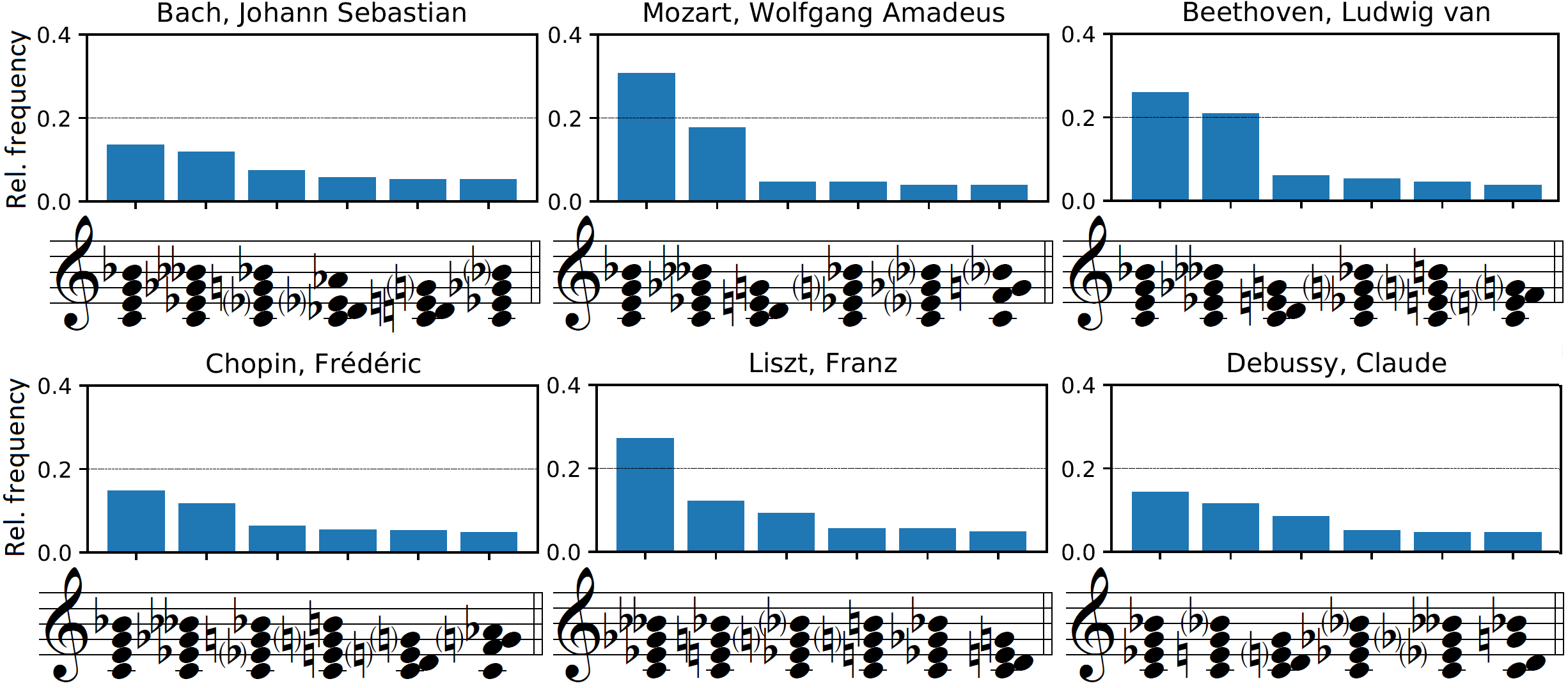}}
  \caption{Tetrachord distribution of six composers of the curated GP dataset.}
  \label{fig:selected_composers_chords_4}
\end{figure}

\subsection{Interval Distribution}
 An interval is the pitch difference between two notes. Intervals can be either melodic intervals or harmonic intervals. A harmonic interval is the pitch difference of two notes being played at the same time. A melodic interval is the pitch difference between two successive notes. We consider both harmonic intervals and melodic intervals as intervals. We calculate the distribution of intervals of six composers. Notes are represented as a list of events in a MIDI format. We calculate an interval as:
\begin{equation} \label{eq:interval}
\Delta = y_{n} - y_{n-1},
\end{equation}
\noindent where $ y_{n} $ is the MIDI pitch of a note and $ n $ is the index of the note. 

We calculate ordered intervals including both positive intervals and negative intervals. For example, the interval $ \Delta $ for an upward progress from $ \text{C}_{4} $ to $ \text{D}_{4} $ is $ 2 $, and the interval $ \Delta $ for a downward progress from $ \text{C}_{4} $ to $ \text{A}_{3} $ is $ -3 $. We only consider intervals between -11 to 11 (included) and discard the intervals outside this region. For example, the value 11 indicates a major seventh interval. Fig. \ref{fig:selected_composers_intervals} shows the interval distribution of six composers. All composers used major second and minor third most in their works. The interval distribution is not symmetric to the origin. For example, J. S. Bach and Mozart used more downward major second than the upward major second. In the works of J. S. Bach, the dip in the interval 0 indicates that repeated notes are less commonly used than non-repeated notes. Other composers used more repeated notes than J. S. Bach. Major seventh and tritone are least used by all composers. Some Romantic and later composers, including Chopin, Liszt, and Debussy used all intervals more uniformly than J. S. Bach from Baroque era.

\begin{figure}[t]
  \centering
  \centerline{\includegraphics[width=\columnwidth]{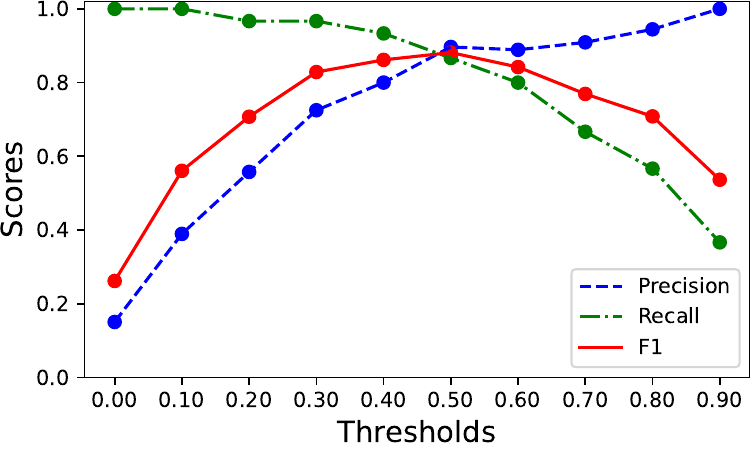}}
  \caption{Precision, recall, and F1 score of solo piano detection.}
  \label{fig:piano_solo_p_r_f1}
\end{figure}

\subsection{Trichord Distribution}
We adopt the set musical theory \citep{forte1973structure} to analyse the chord distribution in GiantMIDI-Piano. A trichord is a set of any three pitch-classes \citep{forte1973structure}. Since GiantMIDI-Piano is transcribed from real recordings, notes of a chord are usually not played simultaneously. We consider notes within a window of 50 ms as a chord. The windows are non-overlapped. Each note only belongs to one chord. For a special case of a set of onsets at 0, 25, 50, 75, and 100 ms, our system first searches chords in a window starting from 0 ms and returns \{0, 25, 50\}. Then, the system searches chords in a window starting from 75 ms and returns \{75, 100\}. We discard the pitch sets with more or less than three notes within a 50 ms window to ensure chords are trichords. The sliding window for counting pitch sets will ensure that there are no overlapped when counting trichords. A major triad can be written as $ \{ 0, 4, 7 \} $, where the interval between 0 and 4 is a major third, and the interval between 4 and 7 is a minor third. We transpose all chords to chords with lower notes $ \text{C} $. For example, a chord $ \{2, 6, 9\} $ is transposed into $\{0, 4, 7\}$. We merge chords with the same prime form. Fig. \ref{fig:selected_composers_chords_3} shows the trichord distribution of six composers. All composers used major triad $ \{0, 4, 7\} $ most followed by minor triad $\{0, 3, 7\}$ in their works. Liszt used more augmented triad $ \{0, 4, 8\} $ than other composers. Debussy used more $ \{0, 2, 5\} $, $ \{0, 2, 4\} $ than other composers which distinguished him from other composers. Fig. \ref{fig:selected_composers_chords_3} shows that composers from different eras have different preferences for using trichords.

\begin{figure*}[t]
  \centering
  \centerline{\includegraphics[width=\textwidth]{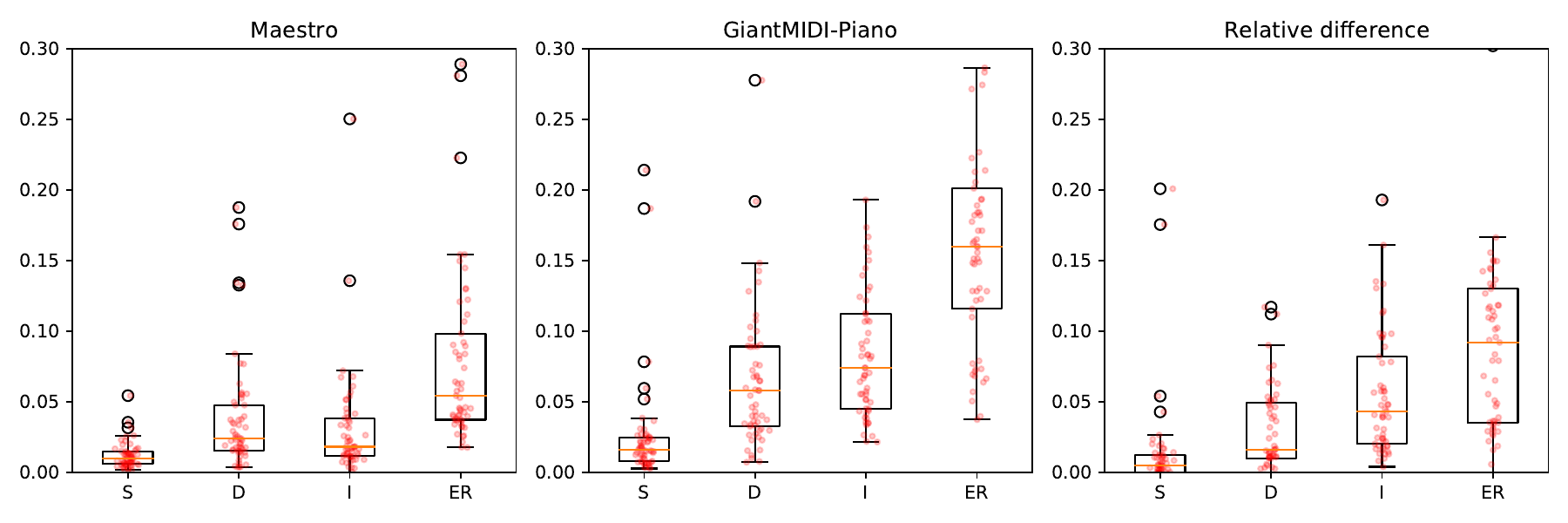}}
  \caption{From left to right: ER of 52 solo piano works in the MAESTRO dataset; ER of 52 solo piano works in the GiantMIDI-Piano dataset; Relative ER between the MAESTRO and the GiantMIDI-Piano dataset.}
  \label{fig:box_plot}
\end{figure*}

\subsection{Tetrachord Distribution}
A tetrachord is a set of any four pitch-classes \citep{forte1973structure}. Similar to trichord, we consider notes within a window of 50 ms as a chord. We discard the pitch sets with more or less than four notes within a 50 ms window to ensure chords are tetrachords. A dominant seventh chord can be denoted as $ \{0, 4, 7, 10\} $. Fig. \ref{fig:selected_composers_chords_4} shows the tetrachord distribution of six composers. Seventh chords such as $ \{0, 2, 6, 9\} $ are transposed to root position seventh chords $ \{ 0, 4, 7, 10 \} $. Fig. \ref{fig:selected_composers_chords_4} shows that Bach, Beethoven and Mozart, and Chopin used dominant seventh $ \{0, 4, 7, 10\} $ most. Liszt used diminished seventh $ \{0, 3, 6, 9\} $ most and Debussy used minor seventh $ \{0, 3, 7, 10\} $ most. J. S. Bach used less dominant seventh compared to the other five composers. The tetrachord distribution of Debussy is different from other composers. Fig. \ref{fig:selected_composers_chords_4} shows that composers from different eras have different preferences for using tetrachords.

\section{Evaluation of GiantMIDI-Piano}

\subsection{Solo Piano Evaluation}
We evaluate the solo piano detection system as follows. We manually label 200 randomly selected music works from 60,724 downloaded audio recordings. We calculate the precision, recall, and F1 scores of the solo piano detection system with different thresholds ranging from 0.1 to 0.9 and show results in Fig. \ref{fig:piano_solo_p_r_f1}. Horizontal and vertical axes show different thresholds and scores, respectively. Fig. \ref{fig:piano_solo_p_r_f1} shows that higher thresholds lead to higher precision but lower recall. When we set the threshold to 0.5, the solo piano detection system achieves a precision, recall, and F1 score of 89.66\%, 86.67\%, and 88.14\%, respectively. In this work, we set the threshold to 0.5 to balance the precision and recall for solo piano detection. 

\subsection{Metadata Evaluation}
We randomly select 200 solo piano works from the full GiantMIDI-Piano dataset and manually check how many audio recordings and metadata are matched. We observe that 174 out of 200 solo piano works are correctly matched, leading to a metadata accuracy of 87\%. Most errors are caused by mismatched composer names. For example, when the keyword $X$ is \textit{Chartier, Mathieu, Nocturne No.1} composed by Chartier, the retrieved YouTube title $ Y $ is \textit{Nocturne No. 1} composed by Chopin. After surname constraint, 136 out of 140 solo piano works are correctly matched, leading to a precision of 97.14\%. We also observe that there are 180 live performances and 20 sequenced MIDI files out of 200 solo piano works.

Furthermore, Table \ref{table:composer_acc} shows the number of matched music works composed by six different composers. \textit{Correct} indicates that the retrieved solo piano works are indeed composed by the composer. \textit{Incorrect} indicates that the retrieved music works are not composed by the composer but are composed by someone else and are attached to the composer. Without surname constraint, Liszt achieves the highest match accuracy of 90\%, while Chopin achieves the lowest match accuracy of 37\%. Table \ref{table:surname_constraint_composer_acc} shows that the match accuracy of Chopin increases from 37\% to 82\% after surname constraint. The accuracy of other composers also increases. The curated GiantMIDI-Piano dataset contains 7,236 MIDI files composed by 1,787 composers. We use a \textit{youtube\_title\_contains\_surname} flag in the metadata file to indicate whether the surname is verified.

\begin{table}[t]
\centering
\caption{Accuracy of retrieved music works of six composers.}
\label{table:composer_acc}
\resizebox{\columnwidth}{!}{%
\begin{tabular}{*{7}{c}}
 \toprule
 & J. S. Bach & Mozart & Beethoven & Chopin & Liszt & Debussy \\
 \midrule
 Correct & 147 & 85 & 82 & 102 & 197 & 29 \\
 Incorrect & 102 & 35 & 70 & 171 & 22 & 9 \\
 \midrule
 Accuracy & 59\% & 71\% & 54\% & 37\% & 90\% & 76\% \\
 \bottomrule
\end{tabular}}
\end{table}

\begin{table}[t]
\centering
\caption{Accuracy of retrieved music works of six composers after surname constraint.}
\label{table:surname_constraint_composer_acc}
\resizebox{\columnwidth}{!}{%
\begin{tabular}{*{7}{c}}
 \toprule
 & J. S. Bach & Mozart & Beethoven & Chopin & Liszt & Debussy \\
 \midrule
 Correct & 129 & 72 & 76 & 96 & 141 & 27 \\
 Incorrect & 44 & 16 & 5 & 21 & 6 & 3 \\
 \midrule
 Accuracy & 75\% & 82\% & 94\% & 82\% & 96\% & 90\% \\
 \bottomrule
\end{tabular}}
\end{table}

\subsection{Piano Transcription Evaluation}

The piano transcription system achieves a state-of-the-art onset F1 score of 96.72\%, onset and offset F1 score of 82.47\%, and an onset, offset, and velocity F1 score of 80.92\% on the test set of the MAESTRO dataset. The sustain pedal transcription system achieves an onset F1 of 91.86\%, and a sustain-pedal onset and offset F1 of 86.58\%. The piano transcription system outperforms the previous onsets and frames system \cite{hawthorne2017onsets, hawthorne2018enabling} with an onset F1 score of 94.80\%.

We evaluate the quality of GiantMIDI-Piano on 52 music works that appear in all of the GiantMIDI-Piano, the MAESTRO, and the Kunsterfuge \citep{kunstderfuge} datasets. Long music works such as Sonatas are split into movements. Repeated music sections are removed. Evaluating GiantMIDI-Piano is a challenging problem because there are no aligned ground-truth MIDI files, so the metrics in \citep{hawthorne2017onsets} are not usable. In this work, we propose to use an alignment metric \citep{nakamura2017performance} called \textit{error rate (ER)} to evaluate the quality of transcribed MIDI files. This metric reflects the substitution, deletion, and insertion between a transcribed MIDI file and a target MIDI file. For a solo piano work, we align a transcribed MIDI file with its sequenced MIDI version using a hidden Markov model (HMM) tool \citep{nakamura2017performance}, where the sequenced MIDI files are from the Kunsterfuge \citep{kunstderfuge} dataset. The ER is defined as the summation of substitution, insertion, and deletion:
\begin{equation} \label{eq:error_rate}
ER = \frac{S + D + I}{N},
\end{equation}
\noindent where $ N $ is the number of reference notes, and $ S $, $ I $, and $ D $ are the number of substitution, insertion, and deletion, respectively. Substitution indicates that some notes are replacements of ground truth notes. Insertion indicates that extra notes are being played. Deletion indicates that some notes are missing. Lower ER indicates better transcription performance.

The ER of music works from GiantMIDI-Piano consists of three parts: 1) performance errors, 2) transcription errors, and 3) alignment errors:
\begin{equation} \label{eq:giantmidi_error}
ER_{\text{G}} = e_{\text{performance}_{\text{G}}} + e_{\text{transcription}_{\text{G}}} + e_{\text{alignment}_{\text{G}}}
\end{equation}
\noindent where the subscript G is the abbreviation for GiantMIDI-Piano. The performance errors $ e_{\text{performance}_{\text{G}}} $ come from that a pianist may accidentally miss or add notes while performing \citep{repp1996art}. The transcription errors $ e_{\text{transcription}_{\text{G}}} $ come from piano transcription system errors. The alignment errors $ e_{\text{alignment}_{\text{G}}} $ come from the sequence alignment algorithm \citep{nakamura2017performance}.

Audio recordings and MIDI files are perfectly aligned in the MAESTRO dataset, so there are no transcription errors. The ER can be written as:
\begin{equation} \label{eq:yamaha_error}
ER_{\text{M}} = e_{\text{performance}_{\text{M}}} + e_{\text{alignment}_{\text{M}}},
\end{equation}
\noindent where the subscript M is the abbreviation for MAESTRO. For a same music work, we assume an approximation $ e_{\text{performance}_{\text{G}}} \approx e_{\text{performance}_{\text{M}}} $ despite the performance among pianists are different. Similarly, we assume an approximation $ e_{\text{alignment}_{\text{G}}} \approx e_{\text{alignment}_{\text{M}}} $ despite the alignment errors are different.

Those approximations are more accurate when the levels of the two pianists are closer. Then, we propose a \textit{relative error} by subtracting (\ref{eq:giantmidi_error}) and (\ref{eq:yamaha_error}):
\begin{equation} \label{eq:relative_error}
r = ER_{\text{G}} - ER_{\text{M}} \approx e_{\text{transcription}_{\text{G}}}.
\end{equation}
\noindent The relative error $ r $ is a rough approximation of the transcription errors $ e_{\text{transcription}_{\text{G}}} $. A lower $ r $ value indicates better transcription quality.

\begin{table}[t]
\centering
\caption{Piano transcription evaluation on the GiantMIDI-Piano dataset}
\label{table:alignment_performance}
\resizebox{\columnwidth}{!}{%
\begin{tabular}{*{5}{c}}
 \toprule
 & D & I & S & ER \\
 \midrule
 Maestro & 0.009 & 0.024 & 0.018 & 0.061 \\
 GiantMIDI-Piano & 0.015 & 0.051 & 0.069 & 0.154 \\
 \midrule
 Relative difference & 0.006 & 0.026 & 0.047 & 0.094 \\
 \bottomrule
\end{tabular}}
\end{table}

Table \ref{table:alignment_performance} shows the alignment performance. The median alignment $ S_{\text{M}} $, $ D_{\text{M}} $, $ I_{\text{M}} $ and $ ER_{\text{M}} $ on the MAESTRO dataset are 0.009, 0.024, 0.021 and 0.061 respectively. The median alignment $ S_{\text{G}} $, $ D_{\text{G}} $, $ I_{\text{G}} $ and $ ER_{\text{G}} $ on the GiantMIDI-Piano dataset are 0.015, 0.051, 0.069 and 0.154 respectively. The relative error $ r $ between MAESTRO and GiantMIDI-Piano is 0.094. The first column of Fig. \ref{fig:box_plot} shows the box plot metrics of MAESTRO. Some outliers are omitted from the figures for better visualization. Some outliers are mostly caused by different interpretations of trills and tremolos. The second column of Fig. \ref{fig:box_plot} shows the box plot metrics of GiantMIDI-Piano. In GiantMIDI-Piano, \textit{Keyboard Sonata in E-Flat Major, Hob. XVI/49} composed by Haydn achieves the lowest ER of 0.037, while \textit{Prelude and Fugue in A-flat major, BWV 862} composed by Bach achieves the highest ER of 0.679 (outlier beyond the plot range). This underperformance is due to the piano is not tuned to a standard pitch with $\text{A}_{4}$ of 440 Hz. The third column of Fig. \ref{fig:box_plot} shows the relative ER between MAESTRO and GiantMIDI-Piano. The relative median scores of S, D, I and ER are 0.006, 0.026, 0.047 and 0.094 respectively. Fig. \ref{fig:box_plot} also shows that there are fewer deletions than insertions.

\section{Conclusion}
We collect and transcribe a large-scale GiantMIDI-Piano dataset containing 38,700,838 transcribed piano notes from 10,855 unique classical piano works composed by 2,786 composers. The total duration of GiantMIDI-Piano is 1,237 hours. The curated subset contains 24,253,495 piano notes from 7,236 works composed by 1,787 composers. GiantMIDI-Piano is transcribed from YouTube audio recordings searched by meta-information from IMSLP.

The solo piano detection system used in GiantMIDI-Piano achieves an F1 score of 88.14\%, and the piano transcription system achieves a relative error rate of 0.094. The limitations of GiantMIDI-Piano include: 1) There are no pitch spellings to distinguish enharmonic notes. 2) GiantMIDI-Piano does not provide beats, time signatures, key signatures, and scores. 2) GiantMIDI-Piano does not disentangle the music score and the expressive performance of pianists.

We have released the source code for acquiring GiantMIDI-Piano. In the future, GiantMIDI-Piano can be used in many research areas, including but not limited to musical analysis, music generation, music information retrieval, and expressive performance analysis.

\section{Acknowledgement}
We thank all anonymous reviewers, editors, and copy editors for their substantial reviews of this manuscript. We thank Prof. Xiaofeng Zhang for passing his composition knowledge to Qiuqiang Kong during his undergraduate study at the South China University of Technology.

\bibliography{TISMIRtemplate}

% For non bibtex users:
%\begin{thebibliography}{citations}
%
%\bibitem {Author:00}
%E. Author.
%``The Title of the Conference Paper,''
%{\it Proceedings of the International Symposium
%on Music Information Retrieval}, pp.~000--111, 2000.
%
%\bibitem{Someone:10}
%A. Someone, B. Someone, and C. Someone.
%``The Title of the Journal Paper,''
%{\it Journal of New Music Research},
%Vol.~A, No.~B, pp.~111--222, 2010.
%
%\bibitem{Someone:04} X. Someone and Y. Someone. {\it Title of the Book},
%    Editorial Acme, Porto, 2012.
%
%\end{thebibliography}

\end{document}